\documentclass[twocolumn,preprintnumbers,amsmath,amssymb,prl]{revtex4-1}

\usepackage{amsmath,amssymb,epsfig,graphicx}

\usepackage{dcolumn}

\usepackage{hyperref}

\def\be{\begin{eqnarray}}
\def\ee{\end{eqnarray}}
\def\bea{\begin{eqnarray}}
\def\eea{\end{eqnarray}}
\newcommand{\Tr}{\mbox{Tr}}

\def\U{\text{U}}
\def\SU{\text{SU}}

\begin{document}

\preprint{}

\title{Comments on Quivers and Fractons
}

\

\author{
Shlomo S. Razamat
}

\

\affiliation{
Department of Physics, Technion, Haifa 32000, Israel\\ 
Institute for Advanced Study, Princeton, NJ 08540, USA
}

\date{\today}

\begin{abstract}

We discuss certain structural analogies between supersymmetric quiver gauge theories and lattice models leading to fracton phases of matter. 
In particular,  classes of quiver models can be viewed as lattice models having sub-system symmetries, dimensions of moduli spaces growing linearly with the 
size of the lattice, and having excitations with limited mobility (with ``excitations'' and ``mobility'' properly defined).  
\end{abstract}


\maketitle


\noindent{\bf Introduction:} 
Lattice models are a very useful effective description of various phases of condensed matter physics. 
There are two ways to understand such lattice models. First, these are quantum mechanical systems with degrees of freedom labeled by lattice points/edges/faces.
As such this is  an example of a $0+1$ quantum field theory (QFT). Second, one literally thinks about the lattice as a spatial arrangement of degrees of freedom, say of atoms. As such the lattice models describe 
excitation propagating in real space and this is the perspective from which the lattice models are usually studied.
(Supersymmetric) quiver gauge theories have a very similar structure. These are QFTs in $(D-1)+1$ dimensions with fields labeled by vertices/edges of a lattice.
The vertices are associated typically  to gauge interactions and the edges to matter fields. One can have also interactions associated to faces of the lattice.
On the other hand one can   try to view the lattice of the quiver theory as a discretization of spatial dimensions. In other words, we can view the quiver theory as a stack, dimension of which is the dimension of the lattice, of $D$-dimensional coupled layers. This approach was pioneered in \cite{ArkaniHamed:2001ca}.

In recent years there is a lot of interest  in the condensed matter literature in a new type of lattices which give rise to the so called {\it fractonic} phases of matter \cite{PhysRevLett.94.040402,PhysRevB.92.235136,Nandkishore_2019}.
These systems have  quite interesting properties. These include, among others, sub-system symmetries, excitations of restricted mobility, and large number of vacua (log of which scales linearly with the size of the lattice). Moreover, these systems pose a challenge to the usual Wilsonian paradigm of going from a lattice theory in the UV to a continuum QFT in the IR \cite{Seiberg:2020bhn}.

In this note we will want to revisit the interpretation of quiver gauge theories as a lattice model and draw some analogies to the lattice theories of fractons. In particular we will consider quiver theories corresponding to a two dimensional periodic (toroidal) lattice.  We will show that when these models are viewed as residing on a spatial lattice they have properties which are characteristic of the fractonic theories. Namely, these models have sub-system symmetries, excitations of restricted mobility and manifolds of vacua (dimension of which scales linearly with the size of the lattice). We will discuss a very particular supersymmetric quiver theory with $D=4$.

\begin{figure}[htbp]
\includegraphics[scale=0.14]{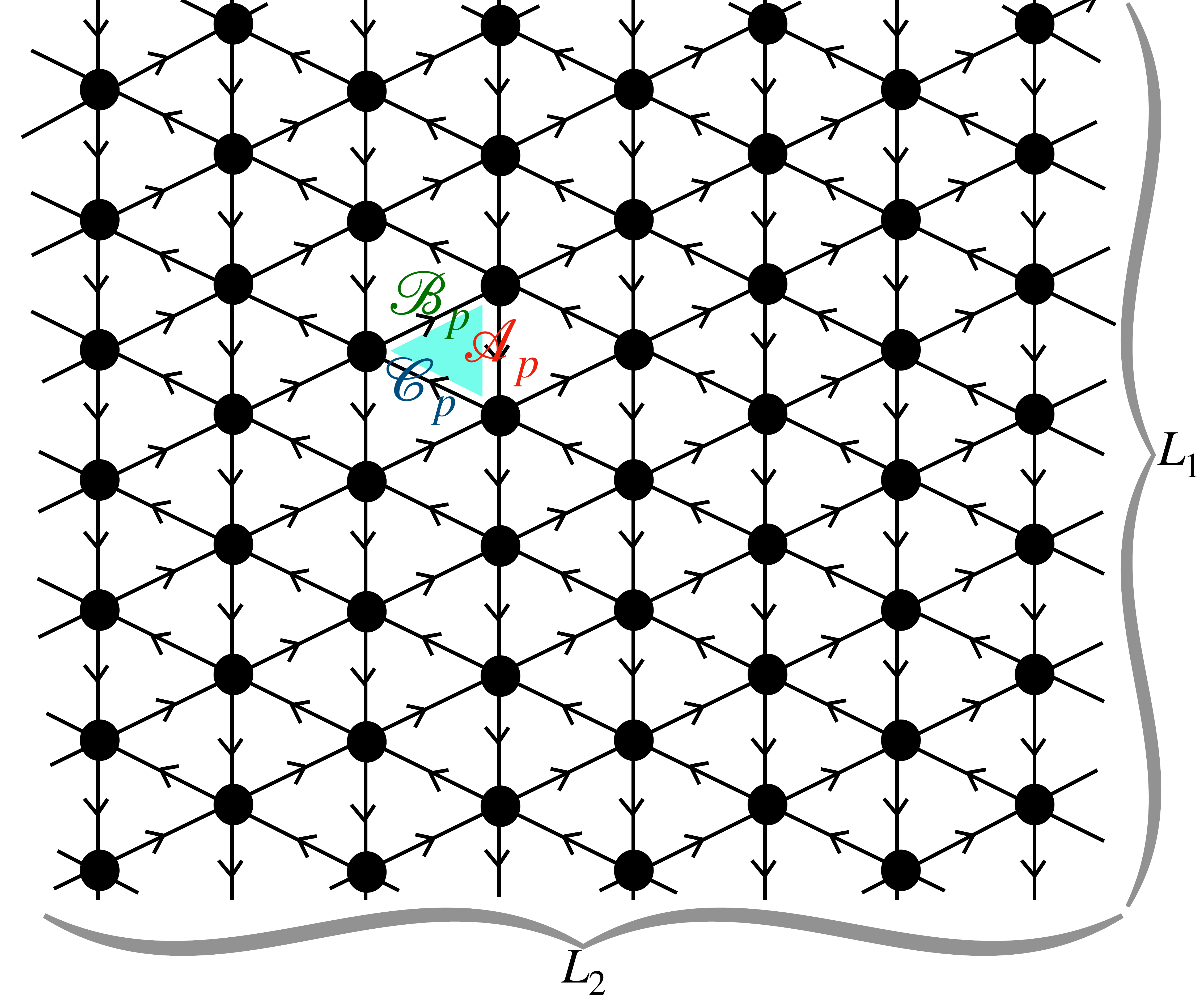}
\caption{An example of a quiver theory. Each node is an $\text{SU}(N)$ gauge group. The lines are chiral superfields in the fundamental representation of the symmetry of the node they point to and anti-fundamental representation of the node they emanate from.}\label{fig:quiver}
\end{figure}

\noindent{\bf A Quiver theory/Lattice:} Let us define an example of a quiver theory. We consider an ${\cal N}=1$ sypersymmetric gauge theory in $D=4$. The gauge group
is $\SU(N)^{L_1\times L_2}$. The matter content consists of a collection of chiral superfields transforming in a fundamental representation of one of the $L_1\times L_2$ $\SU(N)$ gauge factors and in anti-fundamental of another. The matter content can be effectively encoded in a {\it quiver} diagram of Fig. \ref{fig:quiver}.  The lattice has $L_1$ nodes along the vertical axis and $L_2$ nodes along the horizontal axis, and it is periodic. 
The quiver theory has chiral superfields ${\frak A}_p$, ${\frak B}_p$ and ${\frak C}_p$ associated to the three types of edges: the vertical, and the two diagonals.
We can label the fields by one of the two triangular plaquettes, $p$, they reside on.
 The interactions include ${\cal N}=1$ $\SU(N)$ gauge dynamics at each vertex involving the fields ending on that vertex. The fields are in the fundamental representation if the edge   goes into the vertex and anti-fundamental if it exits it. Moreover, there is a superpotential interaction associated to each triangular plaquette,
\be\label{superpotential}
W=\sum_{p}\lambda_p\, \text{Tr}\,{\frak A}_p{\frak B}_p{\frak C}_p\,,
\ee where $p$ labels the plaquette ($\lambda_p$ are  couplings). This quiver theory is conformal meaning that one can continuously turn on certain combinations of the couplings starting from the free theory without breaking conformality and triggering an RG flow, see {\it e.g.} \cite{ArkaniHamed:2001ie,Gaiotto:2015usa}. Such quiver theories appear in numerous contexts and have been extensively studied following \cite{Douglas:1996sw}. The quiver theories are very much reminiscent of layered CS models studied  {\it e.g.} in \cite{CSlattice}. In fact the $D=4$ quiver theory can be thought of as an IR effective description of  a UV $D=6$ $(1,0)$ SCFT compactified on a torus with minimal punctures \cite{Gaiotto:2015usa}. The $D=6$ SCFT
is either the one residing on $N$ M5-branes probing ${\mathbb Z}_{L_1}$ singularity and we have $L_2$ punctures, or ${\mathbb Z}_{L_2}$ singularity and we have $L_1$ punctures.
Taking all couplings to be small one can think of the punctures in either description as ordered along one of the cycles of the torus giving a notion of locality on the lattice.
This is very reminiscent of lattice spin models being an effective description of an underlying UV continuum theory.
Note that for general $L_i$ and $N>3$ there are no supersymmetric marginal or relevant deformations beyond gauge interactions and \eqref{superpotential}, while for $N=3$ we can add marginal baryonic superpotentials to be discussed later, and for $N=2$ we can have relevant mass terms for the fields.

\begin{figure}[htbp]
\includegraphics[scale=0.14]{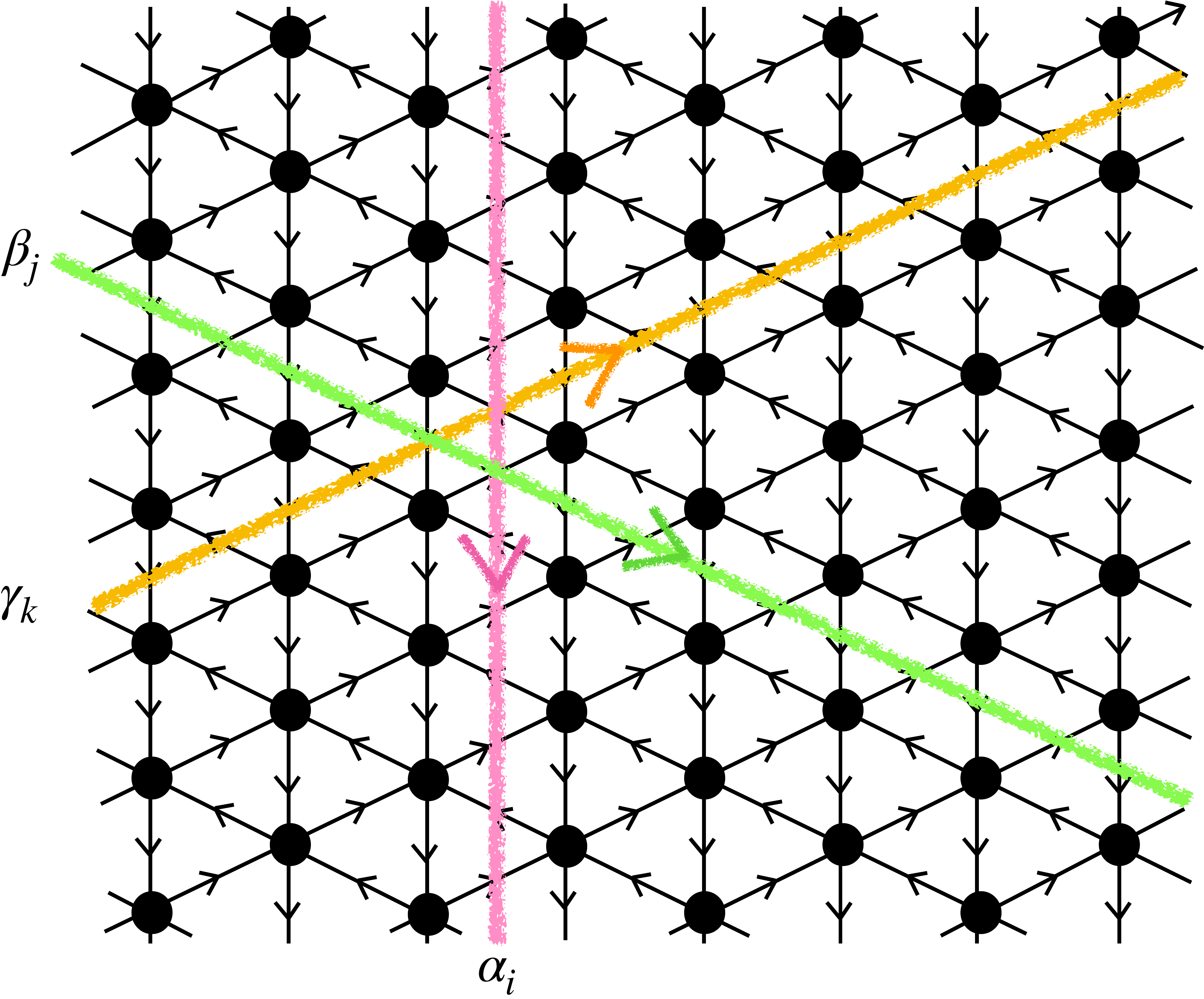}
\caption{Sub-system symmetries.}\label{fig:symm}
\end{figure}

\noindent{\bf Sub-system symmetry:} Let us discuss the global symmetry of the quiver theory.
First, as the theory is conformal in the UV it has an R-symmetry which assigns free charges ($2/3$) to all the chiral superfields.
Moreover, the model possesses additional global symmetries. The dimension of the symmetry group depends on the size of the lattice. We have $L_2$ $\U(1)$ symmetries which 
we label by $\alpha_i$ ($i=1,\cdots, L_2$) associated to the columns of the lattice, see Fig.  \ref{fig:symm}. Under these symmetries 
the ${\frak B}$ fields inside the relevant column have charge $+1$ and the  ${\frak C}$ fields inside the relevant column have charge $-1$. While all ${\frak A}$ fields  and fields outside the column have vanishing charge. This symmetry is thus {\it a sub-system} symmetry of the lattice (This is very reminiscent of the $XY$ plaquette model \cite{PhysRevB.66.054526}.). The model has $L_1$ $\U(1)$ symmetries which 
we label by $\beta_i$ ($i=1,\cdots, L_1$) associated to the diagonals winding to the left and down of the lattice, see Fig.  \ref{fig:symm}. The insistence on having these symmetries determines the gluing of the lattice into a torus. In addition to these  we also have
 $\text{GCD}(L_2,L_1)$ symmetries which 
we label by $\gamma_i$ ($i=1,\cdots, \text{GCD}(L_1,L_2)$) associated to the diagonals winding to the left and up of the lattice.
The ${\frak A}$ fields have charge $(-1,-1)$ under the $(\beta,\gamma)$ symmetries diagonals of which cross it. The fields ${\frak B}$ have charge $+1$ under the $\beta$ symmetry which crosses it and are not charged under the $\gamma$ symmetries. Finally, the fields ${\frak C}$ have charge $+1$ under the $\gamma$ symmetry which crosses it and are not charged under the $\beta$ symmetries. These symmetries (excluding an overall diagonal combination which is redundant) are non-anomalous and consistent with the superpotential \eqref{superpotential}.
 The $\alpha$, $\beta$ and $\gamma$ symmetries (excluding the diagonal combination of each) are thus also {\it sub-system} symmetries.
To summarize the continuous symmetry group of the theory can be written as,
\be
&&G=\U(1)_R\times \frac{\U(1)_{\alpha_i}^{L_2}\times \U(1)_{\beta_i}^{L_2}\times \U(1)_{\gamma_i}^{\text{GCD}(L_1,L_2)}}{\U(1)}\,.\;\;
\ee 
It is useful to define three symmetries $\U(1)_\alpha\times \U(1)_\beta\times \U(1)_\gamma$ such that each is a diagonal combination of the symmetries with the $i$ indices.
The $\U(1)$ projection is such that the diagonal combination of these is factored out. 
One can study more general, twisted, boundary conditions leading to smaller symmetry groups \cite{Hanany:1998it}.


\noindent{\bf Excitations:} Let us consider local gauge invariant excitation on the lattice. By local we mean operators which are constructed from fields of a single edge  on the lattice.
A nice set of such operators is given in terms of the chiral (di)baryons built from the edges of the lattice,
\be
&&A_{ij}= \epsilon^{a_1\dots a_N}\epsilon_{b_1\dots b_N}\prod_{\ell=1}^N \left({\frak A}_{ij}\right)^{b_\ell}_{a_\ell}\,,\nonumber\\
&& B_{ij}= \epsilon_{a_1\dots a_N}\epsilon^{b_1\dots b_N}\prod_{\ell=1}^N \left({\frak B}_{ij}\right)_{b_\ell}^{a_\ell}\,,\\
&&C_{ij}= \epsilon_{a_1\dots a_N}\epsilon^{b_1\dots b_N}\prod_{\ell=1}^N \left({\frak C}_{ij}\right)_{b_\ell}^{a_\ell}\,.\nonumber
\ee  Here we contract the gauge indices and label the fields and the baryons by the symmetries they are charged under. For example, the baryon $A_{ij}$ is charged under $\beta_i$ symmetry and $\gamma_j$ symmetry.  We view these baryons as excitations propagating in the $D$ dimensional continuous space-time as well as on the two dimensional lattice.
Let us  assume first that $\text{GCD}(L_1,L_2)=L_1$ and consider the two-point functions,
\be
\langle B_{ij}(x_1)\left( B_{kl}(x_2)\right)^\dagger\rangle \propto \delta_{ik}\delta_{jl}\,.
\ee The statement on the right-hand side follows simply from symmetry arguments as the baryons at different locations of the lattice are charged under different sub-system symmetries.
We interpret this fact as the baryonic excitations being {\it immobile on the lattice} (though they are free to move in the $D$ dimensional continuous space).
Similar statements can be made about $A$ and $C$ baryons.

\begin{figure}[htbp]
\includegraphics[scale=0.14]{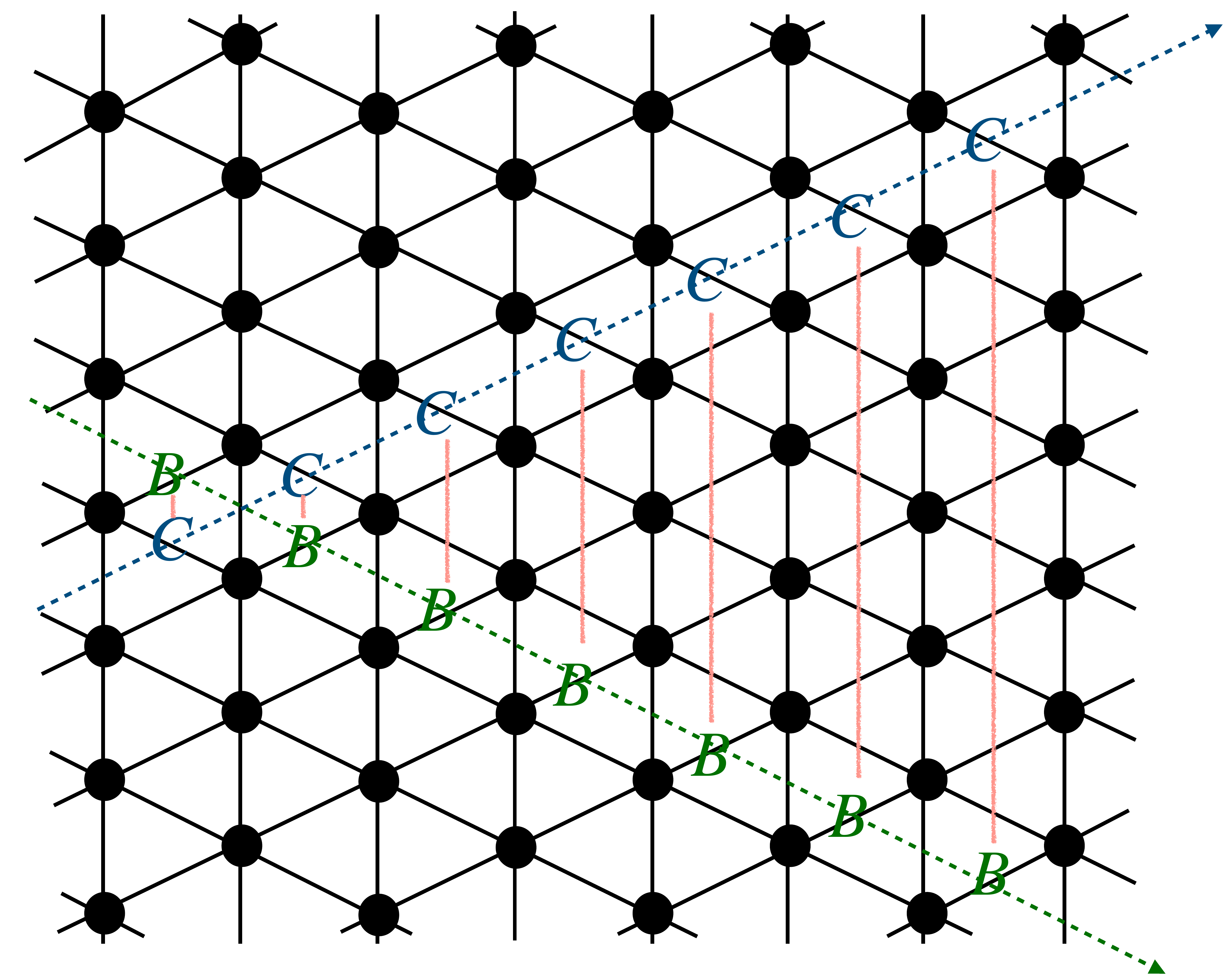}
\caption{Moving baryonic excitations in pairs.}\label{movingpairs}
\end{figure}

On the other hand taking pairs of baryons we have (Here the first index is the $\alpha$ one.),
\be
\biggl< B_{ij}(x_1)C_{ik}(x_2) \left( B_{lm}(x_3)\right)^\dagger \left(C_{ln}(x_4)\right)^\dagger\biggr> \propto \delta_{m-j}\delta_{k-n}\,,\nonumber
\ee as in case the condition on the right-hand side is satisfied the charge under all the symmetries of the correlator is zero. 
 We thus can interpret this result as pairs 
of baryon excitations movable (in principle) along certain directions of the quiver lattice, see Fig. \ref{movingpairs}. One can make analogous statements involving pairs of $B$ and $A$ or $C$ and $A$ baryons.
Finally, if we have a triplet of baryons $A$, $B$, and $C$ residing on the edges of the same plaquette, these can be moved along any direction in a correlated way.

Let us assume now that $\text{GCD}(L_1,L_2)=1$. Then the model has only  $\U(1)^{L_2}_\alpha\times \U(1)_\beta^{L_1}\times \U(1)_\gamma/\U(1)$ symmetry. The baryonic operators $C_{ij}$ can now (in principle) mix with $C_{ik}$, that is with same value of the column index and different values of the diagonal one. In other words the $C$ baryons are movable along the vertical line.
Similar statement holds for the $A$ baryons while the $B$ baryons are still charged under two sub-system  symmetries and thus are immobile on the lattice. 
However, as the symmetry is only broken by  boundary conditions the 
mixing between the $A$ and $C$ baryons at different positions can be only due to a high loop effect order of which scales with the size of the system ($L_i$). Thus assuming $L_i$ are large and the couplings are small (which always can be achieved as the theory is conformal and connected continuously to a free one) the mixing between different sites is strongly suppressed. The excitations are thus effectively immobile even when the symmetries are not there.  The breaking of symmetry is a global effect on the lattice.

Once the couplings are tuned not to be small this argument fails and the baryons acquire mobility. However, tuning the couplings to be finite the notion of locality on the lattice is  lost. For example, the quiver theory we are discussing enjoys an S-duality: a weakly coupled quiver theory is dual to a strongly coupled one with any two of the columns permuted \cite{Gaiotto:2015usa}. Thus in order to have a local interpretation of the lattice we will keep the couplings  small.


\noindent{\bf Vacua:} The quiver theory has a rich structure of  moduli spaces of vacua. First, we have the baryonic branch on which we give vacuum expectation values (vevs) to the baryonic operators. The dimension of this branch scales as $\sim L_1\times L_2$, {\it i.e.} with the number of sites. All the  sub-system symmetries are broken on a generic locus of this branch. One can say that on this branch the excitations with restricted mobility condense.
We will comment on this branch soon but here as we want the baryons to be dynamical excitations we will consider branches on which these do not acquire vevs. 
\begin{figure}[htbp]
\includegraphics[scale=0.14]{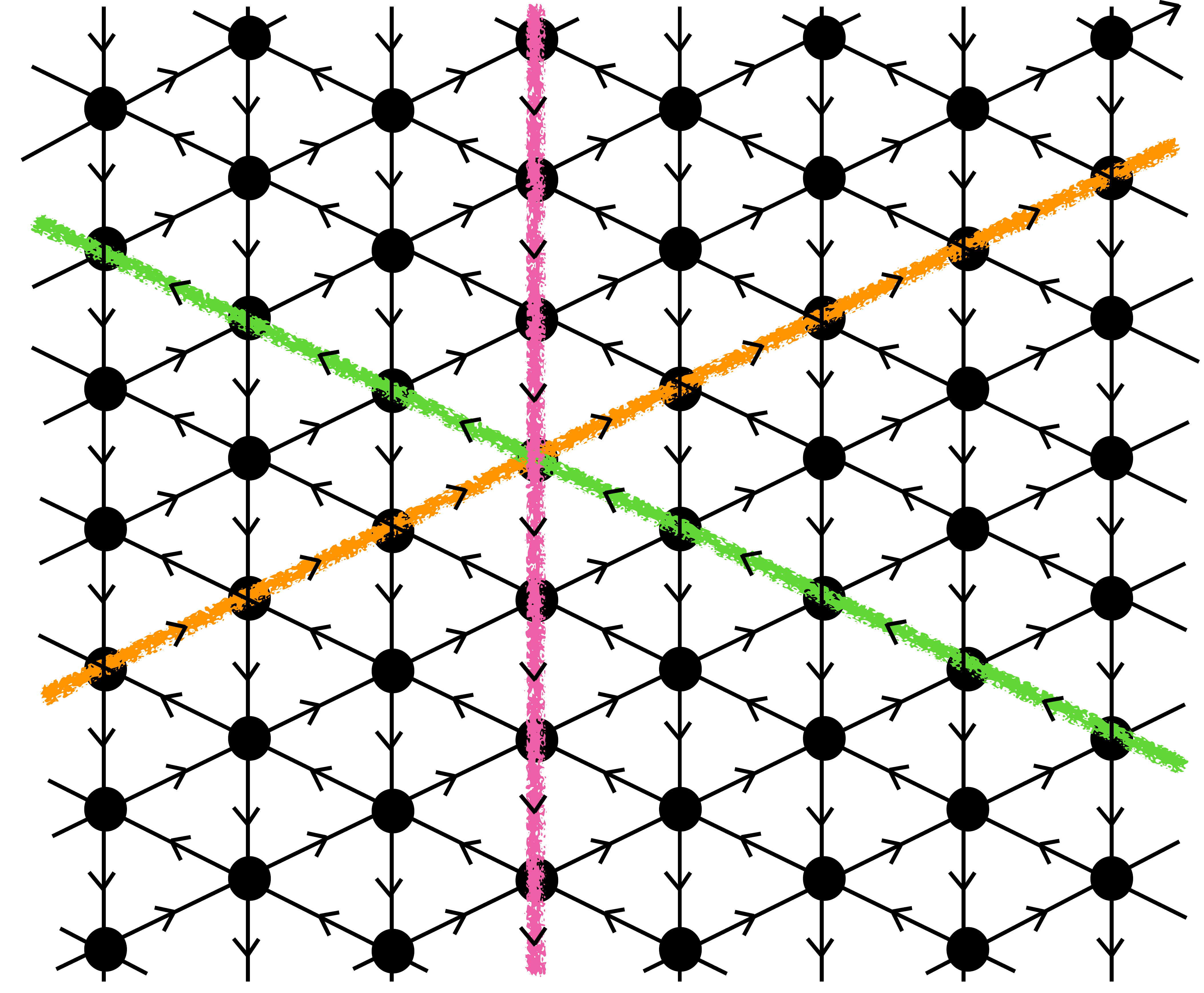}
\caption{The winding operators which can receive a vacuum expectation value and parametrize the moduli spaces of vacua.}\label{fig:vevs}
\end{figure}

We can define three additional natural branches which are parametrized by vevs of operators winding the quiver, see Fig. \ref{fig:vevs}.
First, we have the operators winding vertically along the $\alpha$ cycles,
\be
{\cal A}^{(\ell)}_i =\Tr\,\left(\prod_{p_i} {\frak A}_{p_i} \right)^\ell,\;\; \ell=1\cdots N-1\,,
\ee where $p_i$ are the plaquettes charged under symmetry $\alpha_i$. Note that for the $\text{SU}(N)$ gauge groups we can have independent operators winding up to $N-1$ times. 
 We have $(N-1)\,L_2$ such operators all of which break the same subgroup of the continuous global symmetries. However these have vanishing charges under the sub-system symmetries. 
 We have analogously $(N-1)\,L_1$ operators winding the $\beta$ cycle and $ (N-1)\,\text{GCD}(L_1,L_2)$ operators winding the $\gamma$ cycle. The operators in each one of these branches again have the same charges. The dimension of these branches {\it scales linearly with the size of the lattice}. On any of these branches the gauge groups are broken to an abelian subgroup, the baryons built from the fields getting vevs become free fields and decouple, and all the other fields acquire mass depending on the vev. Thus the theory becomes free in the IR. These branches   can be referred to as generalized Coulomb branch, see {\it e.g.} \cite{Bourton:2020rfo}. In the case of $L_1=1$  the theory has effectively ${\cal N}=2$ supersymmetry with the ${\frak A}$ bifundamentals becoming adjoint (plus a decoupled singlet) chiral superfields, and the operators winding around the $\alpha$ cycle parametrize the familiar ${\cal N}=2$ Coulomb branch.

\

\noindent{\bf A comment on $N=3$:} Taking $N=3$ the baryonic operators become marginal. In particular deforming the Lagrangian by,
\be\label{defN3}
\Delta W_p =\lambda'_p\,\left(A_p+B_p+C_p\right)\,,
\ee  is an exactly marginal deformation ($\lambda'_p$ is the coupling). This is a as the baryons in $\Delta W_p$ are charged under three $\U(1)$s with opposite signs \cite{Green:2010da}.
 This deformation identifies the $\alpha$, $\beta$, and $\gamma$ symmetries under which the plaquette $p$ is charged.
The deformation also lifts part of the baryonic branch of vacua. One can view the plaquettes with the interaction above as impurities which  add directions along which combinations of excitations can move.
Adding densely on the lattice the deformation \eqref{defN3} the excitations  will acquire mobility in various directions.

\

\noindent{\bf The continuum limit:} It is interesting to ask whether we can take a limit such that the two dimensions of the lattice become  continuous.
To do so we might want to introduce a dimensionful parameter and take $L_i$ to infinity. In fact this limit was considered long ago in the framework of \cite{ArkaniHamed:2001ca} in
\cite{ArkaniHamed:2001ie}. See \cite{Hayling:2018fmv} for more recent discussions. There the theory is considered on the baryonic branch with all the sub-system symmetries broken. Performing appropriate scaling of parameters it was argued that the theory is effectively described by the $(1,1)$ little string theory with both $L_i$ large (and the $(2,0)$ SCFT when $L_2$ is large but $L_1=1$) in 6d \cite{Seiberg:1997zk}.
 On the generalized Coulomb branches the sub-system symmetries are preserved and thus one would expect to find unconventional theories of the type discussed in \cite{Seiberg:2020bhn,Seiberg:2019vrp}.


\

\noindent{\bf Summary and Comments:} 
We have drawn a simple analogy between lattice systems leading to fractons and supersymmetric quiver theories. In particular the quiver theories can be naturally associated to a lattice with sub-system symmetries. Operators defined locally on this lattice, such as the baryons, can be thought of as excitations charged under the sub-system symmetry. The problem of studying correlation functions of these operators is analogous to studying the dynamics of the excitations. As the operators are charged under the sub-system symmetries this leads to the excitations  generally having restricted mobility on the lattice. The theory has a large number of vacua, dimension of natural sub-spaces of which, the generalize Coulomb branch, scale linearly with the size of the lattice.
Understanding the continuum limit preserving the sub-system symmetry is related to the question of understanding the dynamics on the generalized Coulomb branch of the quiver theories.

Let us make several comments. First, one can wonder what role did supersymmetry play in our discussion. 
Supersymmetry is important in establishing that the quiver discussed here has moduli spaces of vacua not lifted by quantum effects, as well as the model having
exactly marginal deformations giving us an argument in favor of locality on the lattice.

We can consider giving up conformality in the UV. For example, 
starting from the quiver we have discussed one can construct a huge variety of quiver gauge theories tiling the torus by giving vevs to some (but not all) of the baryons (see {\it e.g.} \cite{Franco:2005rj}). In particular we can preserve say all the $\beta$ and $\gamma$ sub-system symmetries doing so.(These systems and their symmetries were studied {\it e.g.} in \cite{Franco:2015jna} and they have a simple 6d  interpretation \cite{Bah:2017gph}.)
 Such theories will not be conformal in the UV but rather flow to a strongly coupled SCFT in the IR.
 One way to view the generalized Coulomb branches of these models is to start from the quiver of this paper and explore the mixed baryonic and generalized Coulomb branches.
In principle most of what we have discussed applies also to these SCFTs. One interesting question however is whether these theories have limits of their conformal manifold such that the lattice of the quiver is effectively local. Similarly, one can  construct quiver theories with $D<4$ and with more general gauge groups and matter fields, again with the locality on the lattice being an interesting question to be addressed. For example, quiver theories with $D=2$  were discussed in \cite{Kaplan:2002wv,Franco:2016qxh}. As gauge interactions are relevant in lower dimensions we have  more options of constructing UV complete quiver theories.

One of the interesting features of the fractonic systems is the UV/IR mixing: namely the fine details  of the lattice, such as the number of sites, effecting the long distance physics. 
In the quiver example this  is manifest through the dimension of some of the branches of moduli spaces of vacua (as well as the dimension of the conformal manifold and the number of symmetries)  depending on $\text{GCD}(L_1,L_2)$. Moreover,
 the study of quiver theories makes a direct connection to other setups were such UV/IR mixing phenopmena are known, such as the little string theory we mentioned above \cite{SeibergTalk}.
Engineering the quiver theories in string/M-theory ({\it e.g.} studying 6d SCFTs coming from M5-branes probing orbifolds compactified on 2d surfaces with flux in the limit of large flux and large order of the orbifolds \cite{Bah:2017gph,Kim:2018lfo}) might give interesting geometric insights into the continuum limit of the ``fractonic'' (generalized Coulomb) phases of quiver models.

\

\

\noindent{\bf Acknowledgments}:~
We are grateful to Zohar Komargodski, Elli Pomoni, Nati Seiberg, Shu-Heng Shao, and Gabi Zafrir for comments on the draft of the paper.
This research is supported in part by Israel Science Foundation under grant no. 2289/18, by I-CORE  Program of the Planning and Budgeting Committee, by a Grant No. I-1515-303./2019 from the GIF, the German-Israeli Foundation for Scientific Research and Development,  by BSF grant no. 2018204, by the IBM Einstein fellowship  of the Institute of Advanced Study, and by the Ambrose Monell Foundation.

\end{document}